# Nonparametric kernel density estimation of magnitude distribution for the analysis of seismic hazard posed by anthropogenic seismicity

Francis Tong[1] · Stanisław Lasocki[1] · Beata Orlecka-Sikora[1]

**Abstract**
Frequent significant deviations of the observed magnitude distribution of anthropogenic seismicity from the Gutenberg–Richter relation require alternative magnitude–frequency models for probabilistic seismic hazard assessments. Five nonparametric kernel density estimation (KDE) methods are evaluated on simulated samples drawn from four magnitude distribution models: the exponential, concave and convex bi-exponential, and exponential-Gaussian distributions. The studied KDE methods include Silverman's and Scott's rules with Abramson's bandwidth adaptation, two diffusion-based methods (ISJ and diffKDE), and adaptiveKDE, which formulates the bandwidth estimation as an optimization problem. Their performance is assessed for magnitudes from 2 to 6 with sample sizes of 400 to 5000, using the mean integrated square error of cumulative distribution ($\mathrm{MISE}_F$) over 100,000 simulations. Their suitability in hazard assessments is illustrated by the mean of the mean return period (MRP) for a sample size of 1000. Among the tested methods, diffKDE provides the most accurate cumulative distribution function estimates for larger magnitudes. Even when the data are drawn from an exponential distribution, diffKDE performs comparably to maximum likelihood estimation when the sample size is at least 1000. Given that anthropogenic seismicity often deviates from the exponential model, using diffKDE for probabilistic seismic hazard assessments is recommended whenever a sufficient sample size is available.

## Introduction

The earthquake magnitude distribution is an essential component of probabilistic seismic hazard analysis. For this reason, the accurate modeling of this distribution is of paramount importance (Cornell 1968). The most used magnitude distribution model is the exponential distribution. It results from the Gutenberg–Richter magnitude-frequency relation $\log n(M) = a - bM$, where $n(M)$ is the number of earthquakes having magnitudes from the bin centered at $M$, and $a$ and $b$ are constants (Gutenberg and Richter 1944). The probability density function (PDF) of this model reads.

$$f(M) = \begin{cases} 0 & \text{for } M < M_{\min} \\ \beta e^{-\beta(M-M_{\min})} & \text{for } M \geq M_{\min} \end{cases} \quad (1)$$

where $\beta = b\ln 10$, $b$ is the Gutenberg-Richter $b$-value and $M_{\min}$ is the catalog completeness magnitude, defined as the lowest magnitude above which all earthquakes are included in the catalog.

Recent studies of anthropogenic seismicity have shown that the observed magnitude distributions often deviate from exponential models and that the observed distributions may exhibit complex, multimodal structures for which no parametric model can be proposed (Kostoglou et al. 2025, and the references therein). For example, Kostoglou et al. (2025) investigated 63 catalogs of anthropogenic seismicity from the EPISODES platform of the EPOS Thematic Core Service Anthropogenic Hazards (https://episodesplatform.eu; Orlecka-Sikora et al. 2020; Lasocki et al. 2022). The seismicity had been induced by various technological activities, namely reservoir impoundment,

✉ Francis Tong
ftong@igf.edu.pl

Stanisław Lasocki
lasocki@igf.edu.pl

Beata Orlecka-Sikora
orlecka@igf.edu.pl

[1] Institute of Geophysics, Polish Academy of Sciences, Warsaw, Poland

underground mining, conventional and unconventional hydrocarbon extraction, geothermal energy production, and underground gas storage. Goodness-of-fit testing rejected the exponential model for magnitude in 30 cases, and in 16 of which the magnitude distribution was multimodal or had multiple bumps. It was shown that using the exponential distribution might lead to unacceptable inaccuracy of anthropogenic seismic hazard estimates.

Anthropogenic seismicity catalogs are usually big. For example, in hard rock mining at the Pyhäsalmi Mine in Finland, more than 7,000 seismic events were recorded (IS EPOS 2017). Fluid injection for geothermal reservoir stimulation in the Cooper Basin, Australia, resulted in over 27,000 induced earthquakes (IS EPOS 2020). The catalog of seismicity induced by water injection in Oklahoma consists of 22,084 events (IS EPOS 2018). Such considerable sizes of seismicity datasets open an opportunity for using nonparametric statistical techniques. Therefore, as a remedy for the problem of possible critical deviations of anthropogenic seismicity magnitudes from the exponential distribution model, nonparametric kernel density estimation (KDE) (Silverman 1986) was proposed (Kijko et al. 2001; Lasocki and Orlecka-Sikora 2008) as it makes no assumptions about the form of the magnitude distribution.

Within the field of KDE, increasingly advanced implementations are emerging that significantly improve the fit of estimates to the underlying distributions. Therefore, we analyzed five newer implementations to identify the most appropriate KDE for the specific case of magnitude distribution estimation, which could replace currently used algorithms (Lasocki 2021). These were Scott's rule (Scott 1992) and Silverman's rule (Silverman 1986) with Abramson's adaptive bandwidth method (Abramson 1982), the improved Sheather–Jones method (Botev et al. 2010), the diffKDE method (Pelz et al. 2023), and the adaptiveKDE method (Shimazaki and Shinomoto 2010).

We tested the KDE methods on synthetic data generated from three magnitude distribution models: exponential, bi-exponential, and exponential-Gaussian. The KDE estimates were also compared to the maximum likelihood estimation (MLE) estimates obtained when assuming the data follow the Gutenberg–Richter relation; hence, its distribution is exponential.

## Materials and methods

### Kernel density estimation

Kernel density estimation (KDE) is a nonparametric statistical method for estimating probability density functions (PDFs). The kernel density estimator for a sample $x = \{x_1, x_2, \ldots, x_n\}$ is defined as (e.g., Silverman 1986):

$$\hat{f}(x) = \frac{1}{nh} \sum_{i=1}^{n} K\left(\frac{x - x_i}{h}\right) \tag{2}$$

where

$\hat{f}(x)$ is the KDE approximation of the true probability density $f$,

$K(x)$ is the kernel function,

$h$ is the bandwidth.

The kernel function can be any symmetric, nonnegative function that integrates to 1. Since the distribution models considered in this study had semi-infinite supports, we used exclusively the Gaussian kernel,

$$K(x) = \frac{1}{\sqrt{2\pi}} e^{-\frac{1}{2}x^2} \tag{3}$$

As it ensured the infinite support of $\hat{f}(x)$, for this kernel, the KDE estimates of the PDF and cumulative distribution function (CDF) are, respectively:

$$\hat{f}(x) = \frac{1}{nh\sqrt{2\pi}} \sum_{i=1}^{n} \exp\left[-\frac{\left(x - x_i\right)^2}{2h^2}\right] \tag{4}$$

$$\hat{F}(x) = \frac{1}{n} \sum_{i=1}^{n} \Phi\left(\frac{x - x_i}{h}\right) \tag{5}$$

where $\Phi(\xi) = \frac{1}{\sqrt{2\pi}} \int_{-\infty}^{\xi} e^{-t^2/2} \mathrm{d}t$ is the standard normal CDF.

The kernel density estimator is local. When the PDF of the random variable, $X$, is sharply zeroed outside a semi-finite interval $\left[x_*, \infty\right)$, which is the case of magnitude distribution (Eq. 1), then the estimate (Eq. 4) would have a spurious mode close to $x_*$, tending to zero at $x_*$. In such cases, the data sample $x = \{x_1, x_2, \ldots, x_n\}$ is mirrored symmetrically around $x_*$ constructing the new sample $x\prime = \{2x_* - x_n, .., 2x_* - x_2, 2x_* - x_1, x_1, x_2, \ldots x_n\}$ whose KDE estimate (Eq. 4) is $\hat{f}\prime(x)$. The PDF of $X$ is $\hat{f}(x) = 2\hat{f}\prime(x)$.

The bandwidth, $h$, determines how much smoothing is applied to the density, and its choice affects the resultant estimate more than the choice of kernel function. There are many ways to select the bandwidth $h$. The five used in this paper are presented below.

#### Scott's rule

Under the assumption that the data follow a normal distribution, Scott derived an optimal bandwidth as:

$$h = \left(\frac{4}{3}\right)^{\frac{1}{5}} \sigma n^{-1/5} \tag{6}$$

where $\sigma$ is the standard deviation of the data and $n$ is the sample size (Scott 1992).

### Silverman's rule

For data that are close to normal, Silverman proposed the following rule of thumb:

$$h = 0.9\min\left(\sigma, \frac{\mathrm{IQR}}{1.34}\right)n^{-1/5} \tag{7}$$

where $\sigma$ is the standard deviation of the data, IQR is the interquartile range, and $n$ is the sample size (Silverman 1986).

### Adaptive bandwidths to correct Scott's and Silverman's rules

Magnitude distributions are exponential-like. Samples drawn from such distributions are unevenly populated, with sparse data from tails. The Gaussian kernel density estimator (Eq. 4), with a constant bandwidth as in Scott's and Silverman's rules (Eqs. 6, 7), produces spurious irregularities in the intervals where data are sparse (Botev et al. 2010). On the other hand, accurate estimation of the tail of an earthquake magnitude distribution is crucial for accurate estimation of extreme event exceedance probabilities, which is an important part of seismic hazard assessment. The problem due to the sparsity of data in some intervals, e.g., in tails like in the case of earthquake magnitudes, can be alleviated by methods that adapt kernel widths locally at the data points (e.g., Loftsgaarden and Quesenberry 1965; Breiman et al. 1977; Abramson 1982; Silverman 1986; Izenman 1991 and the references therein; Terrell and Scott 1992).

We complete Scott's and Silverman's rules with Abramson's adaptive bandwidth method (Abramson 1982). The bandwidth is adjusted by a factor inversely proportional to the square root of the density estimate. In effect, this reduces the bandwidth where there are many samples and increases it where samples are sparse. In our case, it is implemented as follows:

(1) Compute a pilot density estimate $\widetilde{f}(M_j)$ by convolution of the binned relative frequencies with a Gaussian filter
(2) Calculate the geometric mean $g$ of the pilot density estimates at each point, $\left\{\widetilde{f}(M_j)\right\}_{j=1}^{n}$

The bandwidth for the kernel centered at $M_j$ is then:

$$h_j = h_0\left(\frac{g}{\widetilde{f}}\right)^{-\alpha} \tag{8}$$

where $h_0$ is the bandwidth obtained from a fixed bandwidth selection method (Silverman or Scott).

$g$ is the geometric mean of the pilot density estimates

$\widetilde{f}$ is the pilot density estimate

$\alpha$ is the sensitivity factor

Abramson indicated $\alpha = 0.5$ as an optimal choice.

### Improved Sheather–Jones (ISJ) and DiffKDE methods

Botev et al. (2010) proposed the improved Sheather–Jones (ISJ) bandwidth estimation method, in which they exploited the fact that the Gaussian kernel is a fundamental solution to the heat diffusion partial differential equation (PDE) in $x$ and time $t$

$$\frac{\partial}{\partial t}u(x,t) = \frac{1}{2}\frac{\partial^2}{\partial x^2}u(x,t) \tag{9}$$

Their adaptive kernel density estimation is based on the smoothing properties of the linear diffusion process. The approach of Botev et al. (2010) was further developed by Pelz et al. (2023). Their diffusion-based KDE (diffKDE) allows for adjusting the diffusion intensity (adaptive smoothing) in space.

### AdaptiveKDE

Another way to create an adaptive bandwidth estimator is to reformulate it as an optimization problem. Shimazaki and Shinomoto (2010) proposed minimizing the L2 loss function between the kernel estimate and the true density function. Unlike Abramson's method, which varies the bandwidth only at sample points, this method returns variable bandwidth values across the domain where the data are provided. Performing successive iterations of optimization within each local interval will return an optimal bandwidth for each interval.

### Synthetic catalogs

The performance of the five KDE methods was tested on synthetic magnitude distributions generated by sampling the PDF of each distribution (Eqs. 1, 11, and 12). Sampling was done using the inverse transform method, and $10^5$ simulations were performed for each model distribution. The catalog data were drawn from the following distributions.

### Exponential distribution

The exponential distribution model (Eq. 1) results in the Gutenberg–Richter magnitude–frequency relation (Gutenberg and Richter 1944). We studied this model with the parameters $b = 0.7, 1.0, 1.3$, and $M_{\min} = 0.5$ (Fig. 1).

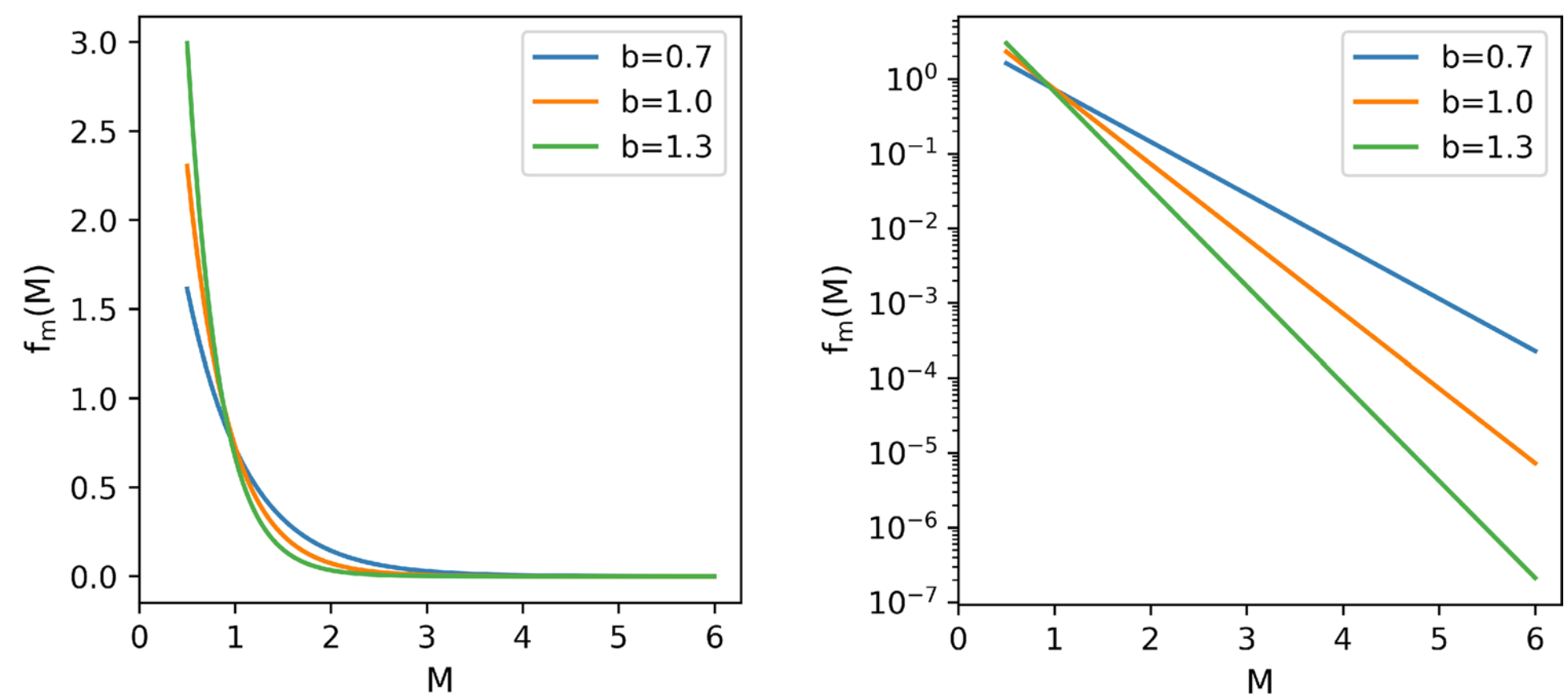

**Fig. 1** Exponential distribution models used to generate synthetic catalogs. The figure on the left uses a linear scale for the vertical axis, whereas the figure on the right uses a logarithmic scale

## Bi-exponential distribution

The bi-exponential distribution results from the bi-linear frequency–magnitude relation

$$\log n(M) = \begin{cases} a_1 - b_1 M & \text{for } M \le M_t \\ a_2 - b_2 M & \text{for } M > M_t \end{cases} \tag{10}$$

(Utsu 1999) where $M_t$ is the transition magnitude between the two different exponential distributions. This is a deviation from the Gutenberg–Richter distribution in that there are two exponential distributions with two different b-values instead of a single exponential distribution with one b-value. Its convex form, $b_1 < b_2$, models a change of the frequency distribution from small to comparatively larger magnitude events resulting from the finite thickness of the seismogenic crust, which discriminates against larger magnitudes. (e.g., Pacheco et al. 1992; Okal and Romanowicz 1994; Sornette and Sornette 1999). However, in anthropogenic seismicity, there is observational evidence of both convex and concave bi-linearity (e.g., Kijko et al. 1987; Johnston and Einstein 1990; Baig and Urbancic 2013; Maghsoudi et al. 2014, 2016).

The PDF following Eq. 10 reads.

$$f(M) = \begin{cases} \lambda\beta_1 e^{-\beta_1(M-M_{\min})} & forM \le M_t \\ \mu\beta_2 e^{-\beta_2(M-M_{\min})} & forM > M_t \end{cases} \tag{11}$$

where $\lambda = \left(1 - \left(1 - \frac{\beta_1}{\beta_2}\right)e^{-\beta_1(M_t - M_{\min})}\right)^{-1}$, $\mu = \lambda \frac{\beta_1}{\beta_2} \frac{e^{-\beta_2(M_t - M_{\min})}}{e^{-\beta_1(M_t - M_{\min})}}$
.

We used the bi-exponential model with the parameters $b_1 \in [0.7, 1.3]$, $b_2 = 2 - b_1$, $b = \beta \ln 10$, $M_{\min} = 0.5$, $M_t = 2.0$ (Fig. 2).

## Exponential-Gaussian distribution

For the last type of distribution used to generate synthetic data, we used a combination of exponential and Gaussian distributions. This is a deviation from the Gutenberg–Richter distribution in that on top of an exponential distribution there is a Gaussian distribution centered at $M_t = 3.0$ which we call the transition magnitude. This combined type of distribution can be used to model the effect of characteristic earthquakes hypothesized and observed in tectonic seismicity (e.g., Wesnousky et al. 1983; Schwartz and Coppersmith 1984; Jackson and Kagan 2011; Parsons et al. 2018), but also observed in

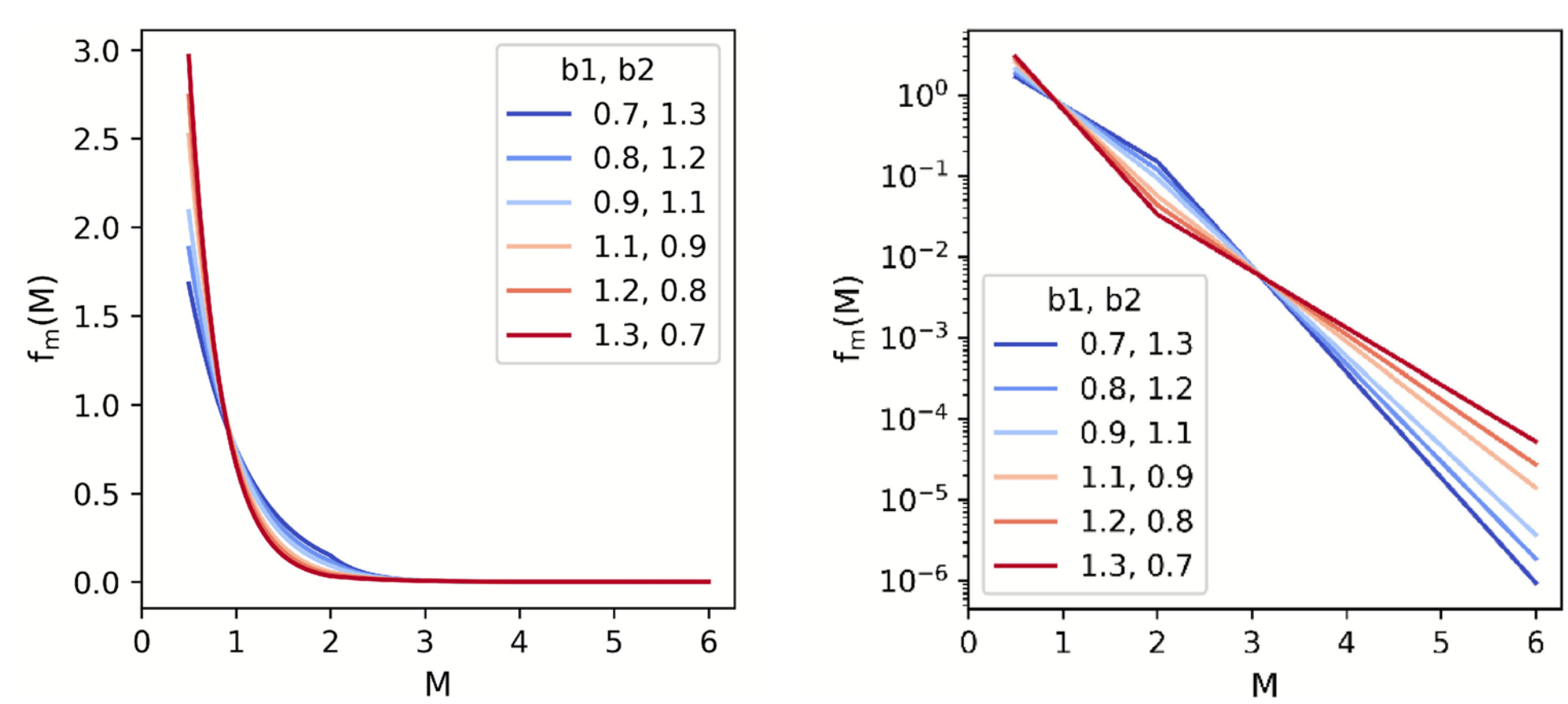

**Fig. 2** Bi-exponential distribution models used to generate synthetic catalogs. The figure on the left uses a linear scale for the vertical axis, whereas the figure on the right uses a logarithmic scale

anthropogenic seismicity (e.g., Eaton et al. 2014; Igonin et al. 2018). The model PDF reads

$$f(M) = p\beta e^{-\beta(M-M_{\min})} + \frac{1-p}{\sqrt{2\pi}\sigma} e^{-\frac{(M-M_t)^2}{2\sigma^2}} \quad (12)$$

We used this model with the parameters $b = \beta \ln 10$ for $b = 1.0, M_{\min} = 0.5, M_t = 3.0,\ \sigma = 0.3,\ p = 0.85, 0.9,$ and 0.95 (Fig. 3).

## Simulation experiments

Between 50 and 5000 samples were generated from each probability distribution. This was repeated to obtain $10^5$ independent simulation experiments for each distribution. We were looking for solutions primarily for human-induced seismicity. Depending on the inducing technological processes, the maximum size of anthropogenic seismic events is between magnitudes 4 and 6 (Lasocki and Orlecka-Sikora 2021). Therefore, the samples and subsequent analysis were restricted to the range beginning at $M_{\min} = 0.5$ and ending at $M = 6$.

The five adaptive KDE methods were applied for each simulation to obtain an estimated PDF $\hat{f}$ for each method for each simulation. Each PDF, $\hat{f}$, was then used to obtain the cumulative distribution function (CDF) estimates, $\hat{F}$, by integration:

$$\hat{F}(x) = \int_{-\infty}^{x} \hat{f}(t)\mathrm{d}t \quad (13)$$

The most widely used metrics of the global accuracy of the density estimator are the mean integrated square error, MISE defined by

$$\mathrm{MISE}\left(\hat{f}\right) = \mathbf{E}\left[\int \left(\hat{f}(x) - f(x)\right)^2 \mathrm{d}x\right], \quad (14)$$

where $f$ is the true density and $\hat{f}$ is its estimator. However, the target of this research was to select the KDE method that was most appropriate for the probabilistic seismic hazard analysis, which requires accurate estimates of the CDF of magnitude. Furthermore, for obvious reasons, the accuracy of CDF estimates for larger magnitudes is of paramount importance. Therefore, the suitability of the tested KDE methods for hazard analysis purposes was determined by the global accuracy of the CDF estimator rather than the density estimator. This accuracy was parameterized by the mean integrated squared error of CDF, $\mathrm{MISE}_F$, defined by:

$$\mathrm{MISE}_F(\mathrm{KDE}) = \mathbf{E}\left[\int \left(\hat{F}(M) - F(M)\right)^2 \mathrm{d}M\right], \quad (15)$$

where $F$ was the true CDF of the distribution from which the samples were drawn and $\hat{F}$ was the CDF obtained through estimation. The performance of each KDE method was settled by the $\mathrm{MISE}_F$ over all $10^5$ simulations. Smaller $\mathrm{MISE}_F(\mathrm{KDE})$ signified a better KDE estimation method. We also fit the exponential model (Eq. 1) to the simulated samples using the maximum likelihood estimation (MLE) method and calculated related $\mathrm{MISE}_F(\mathrm{EXP})$.

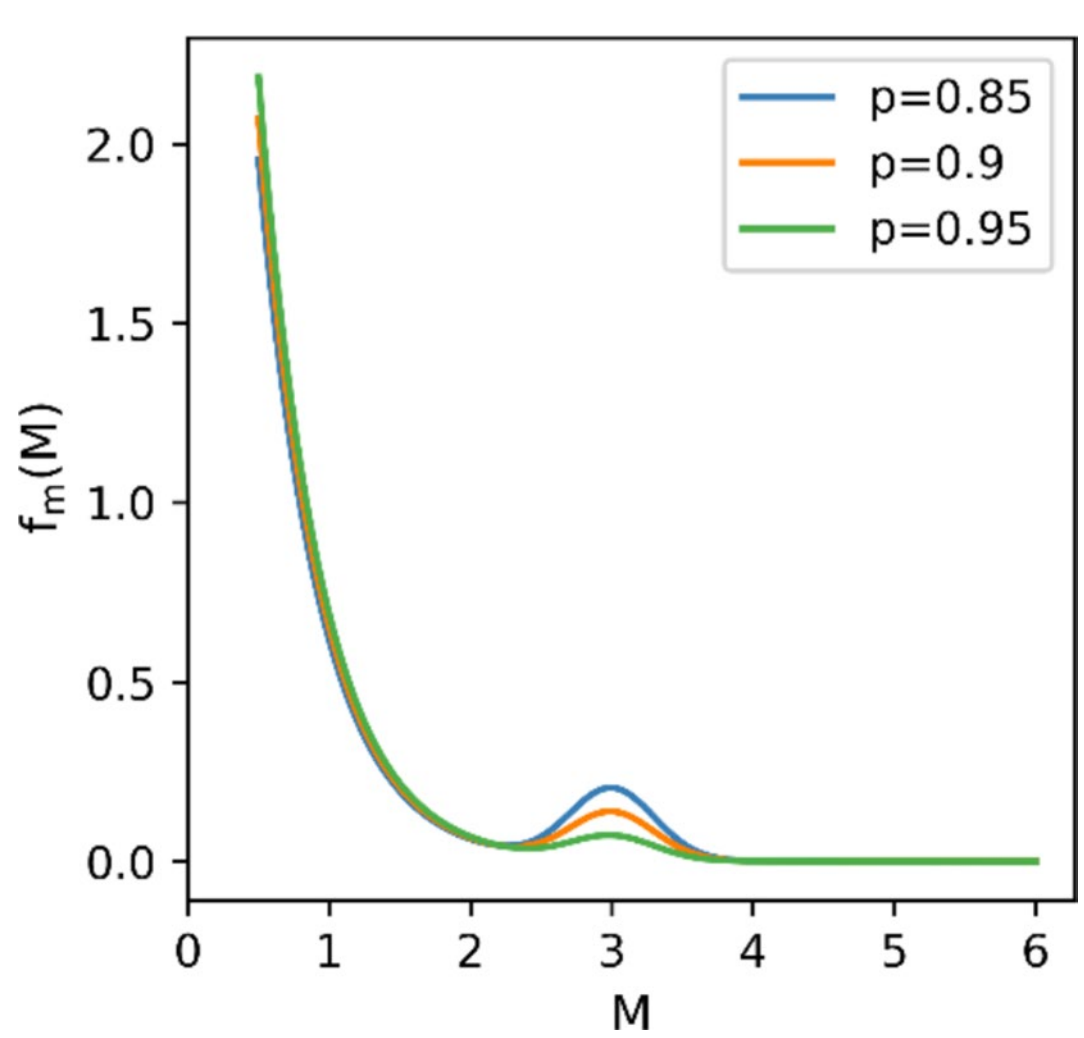

**Fig. 3** Exponential-Gaussian distribution models used to generate synthetic catalogs

Earthquake magnitude distribution is essential for assessing seismic hazard. The mean return period (MRP) is one of the most widely used parameters of seismic hazard as it is the average time between successive earthquakes with magnitude $M$ or higher, assuming that the earthquake-generating process is Poissonian:

$$\mathrm{MRP}(M) = \frac{1}{\lambda(1 - F(M))} \quad (16)$$

where $\lambda$ is the activity rate of events with a magnitude greater than or equal to the magnitude of completeness, and *F(M)* is the CDF of these events.

The performance of the studied KDEs in seismic hazard assessment was investigated. To this end, for each KDE and each model of the underlying probability distribution, mean CDF estimates were calculated for a given magnitude range. The MRP computed using these mean estimates was then compared with the MRP calculated by applying the exact model CDFs. Lambda was assumed to be 10 events per day.

## Results and discussion

We were looking for solutions applicable to seismic hazard analysis that focus on larger magnitude events that can cause damage. For this reason, particular attention was paid to the accuracy of the estimates of both CDF and MRP for larger magnitudes. Therefore, although synthetic catalogs were generated starting from magnitude $M_{\min} = 0.5$, we present here the results concerning magnitudes at and above 2 for a sample size of 400 or more. These sample sizes ensured a non-negligible probability of a magnitude $\geq 2.0$ occurrence in the generated samples. Results across the full magnitude range and all sample sizes are provided in Appendix.

## Exponential distribution

The results when the underlying distribution is exponential (Eq.1) are shown in Fig. 4–the accuracy of estimations expressed in terms of $\mathrm{MISE}_F$ (Eq.15), and Fig. 5–the MRP (Eq.16). When exact parametric data models are known, MLE outperforms nonparametric estimation methods. Therefore, it is not surprising that when an exponential distribution underlay the data, in Fig. 4, $\mathrm{MISE}_F(\mathrm{EXP}) < \mathrm{MISE}_F(\mathrm{KDE})$ for every KDE method. However, all $\mathrm{MISE}_F$ values of the KDE methods were small. The largest $\mathrm{MISE}_F$ values occurred when the sample size was the smallest and $b = 0.7$–about $10^{-5}$ for MLE and $4 \times 10^{-5}$ for Scott's rule (the KDE method with the largest $\mathrm{MISE}_F$ value observed).

Among the KDE methods, diffKDE and ISJ had nearly the same performance. The performance of Silverman's method was similar to those three mentioned when $b$ was the smallest ($b = 0.7$), that is, when the range of larger magnitudes was best populated. For other values of $b$, the performance of Silverman's method worsened. As the sample size increased, the accuracy of estimation of all studied kernel methods increased, which was expressed by the decrease in $\mathrm{MISE}_F$.

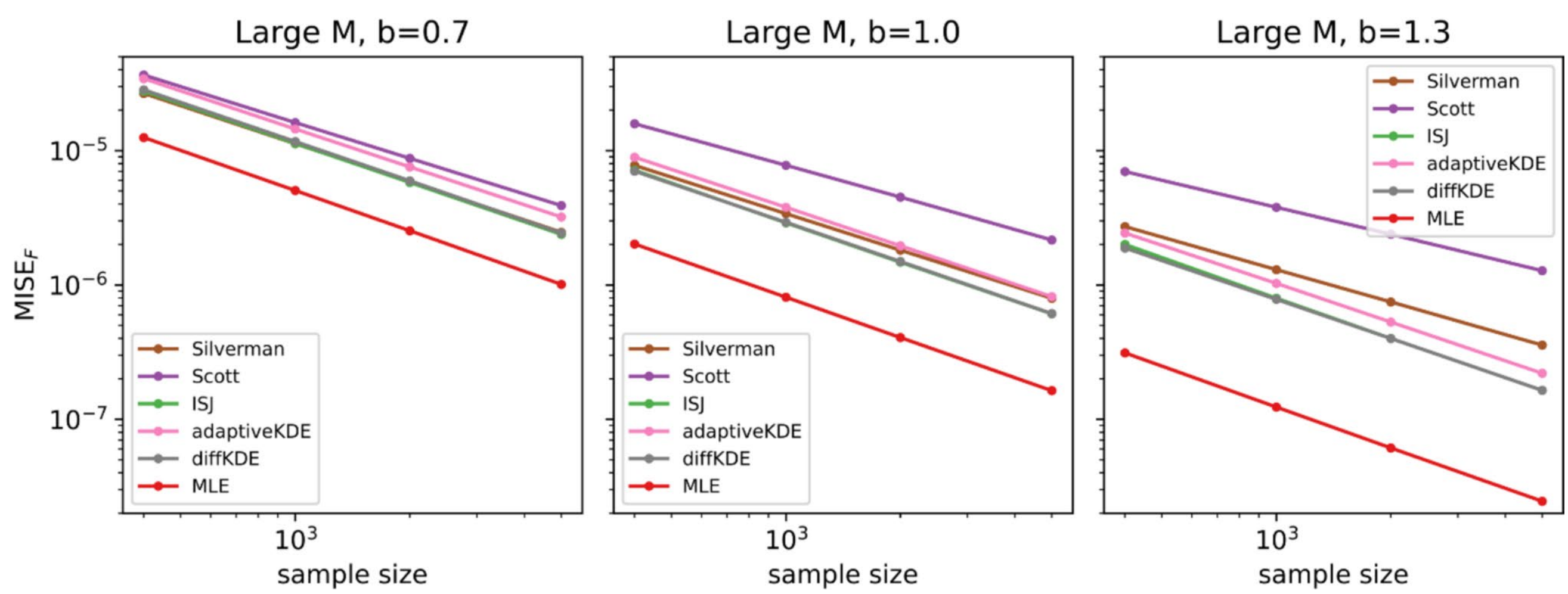

**Fig. 4** $\mathrm{MISE}_F$ results for the data drawn from the exponential distribution (Eq. 1) with $b = 0.7$–left, $b = 1.0$–middle, and $b = 1.3$–right. Colors mark the KDE methods of CDF estimation and the MLE estimate of CDF assuming the exponential distribution model (red)

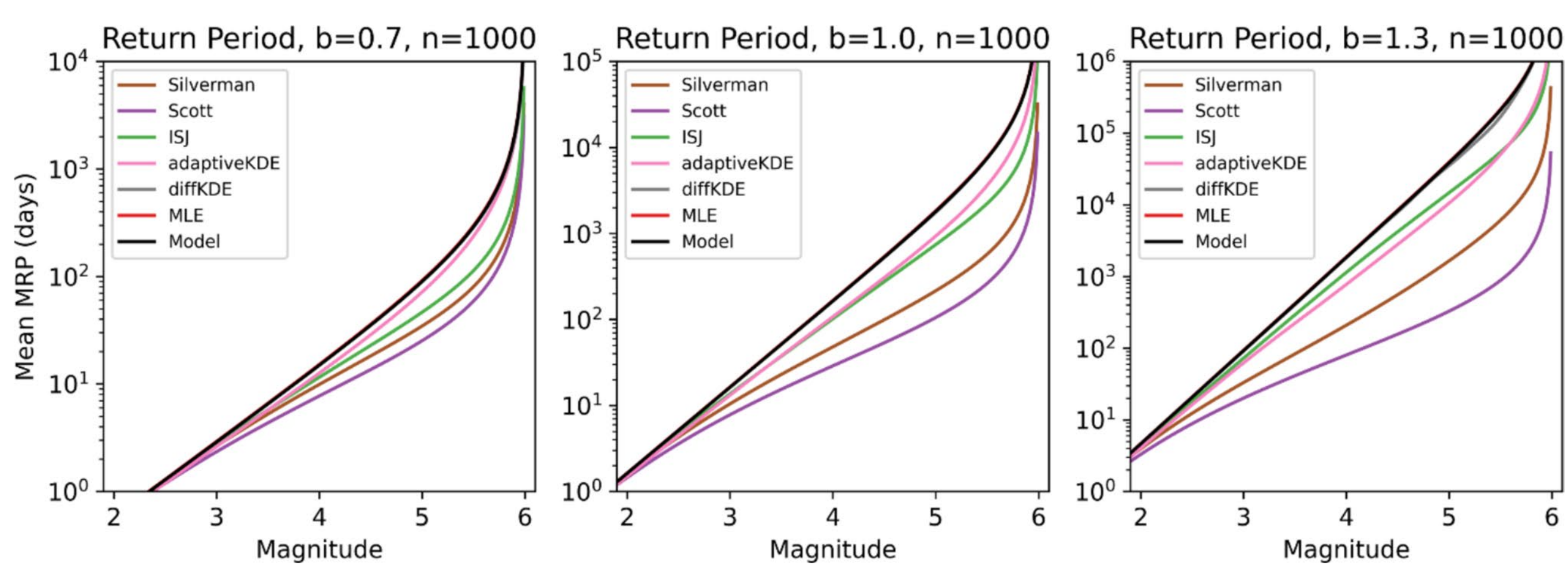

**Fig. 5** MRP for the 1000 magnitude values drawn from the exponential distribution (Eq. 1) with $b = 0.7$–left, $b = 1.0$–middle, and $b = 1.3$–right. Colors mark MRPs for CDFs obtained from the studied KDE methods, for CDF from the MLE assuming the exponential distribution model (MLE—red), and for the CDF of data distribution (Model—black)

The MRP based on the CDF estimate from MLE and the CDF of the model must be nearly the same–discrepancies may result only from the finiteness of the sample used in the MLE estimation. Therefore, the MRP plot in Fig. 5 shows the black line from the model completely overlapping the red line from MLE. However, it is worth noting that the MRP based on diffKDE (the gray line) is nearly identical to the previous two. A slight discrepancy appears only for the simulated data with $b = 1.3$ in the largest magnitude range between 5 and 6. This result suggests that when a sample is suitably big (here, 1000), diffKDE can be safely used in hazard analysis even if the magnitude data follow the Gutenberg-Richter relation and the exponential distribution model.

$\mathrm{MISE}_F$ results from integration over the whole range of M (Eq. 15), whereas the computation of MRP is based on the mean of CDF estimates for individual M from this range. This difference in the ways used to obtain $\mathrm{MISE}_F$ and MRP may lead to varying assessments of method performance using $\mathrm{MISE}_F$ and MRP metrics. Such a case is seen in Figs. 4 and 5. Although the $\mathrm{MISE}_F$ of diffKDE and ISJ are nearly indistinguishable (Fig. 4), the MRPs of these two KDEs are distinct from each other (Fig. 5). The MRP of diffKDE (gray) is coincident with the MRP computed using the exact CDF of the model (black), while ISJ (green) provides a distinctly lower MRP. Therefore, since sole $\mathrm{MISE}_F$ comparisons may not correctly report the consequences for hazard estimates, both $\mathrm{MISE}_F$ and the mean MRP should be considered.

## Bi-exponential distribution

In the following experiment, data were drawn from non-exponential distributions with increasing deviations from exponentiality. These deviations were not pronounced and can be found in actual observations of anthropogenic seismicity (Kostoglou et al. 2025). The $\mathrm{MISE}_F$ (Eq.15) results when the data distribution is bi-exponential (Eqs.10 and 11) are shown in Fig. 6–concave bi-exponential and Fig. 7–convex bi-exponential. The MRPs (Eq.16) for the studied bi-exponential distributions are shown in Fig. 8–concave and Fig. 9–convex.

When considering the $\mathrm{MISE}_F$, diffKDE and ISJ outperformed the other KDE methods (Figs. 6 and 7). Their performance practically did not change when the data distribution deviated more from exponentiality ($\mathrm{abs}(b_1 - b_2)$ increased). As sample size increased, the $\mathrm{MISE}_F$ values of all KDE methods decreased at more or less the same rate. In the case of mean MRP, the diffKDE method provided the most accurate estimates out of all the KDE methods (Figs. 8 and 9). In fact, these diffKDE estimates are visually indistinguishable from the actual (model) MRPs. The diffKDE correctly estimated this seismic hazard parameter.

The distribution from which the data were drawn was not exponential. To illustrate how much the use of the incorrect distribution model might affect hazard estimates, we also computed $\mathrm{MISE}_F$ and MRP using the MLE method with the exponential distribution model. Obviously, $\mathrm{MISE}_F$ values of the MLE method are greater than the $\mathrm{MISE}_F$ values of the KDE methods when the sample size is big enough (Figs. 6 and 7). As it is seen in Figs. 8 and 9, applying MLE under the incorrect assumption of an exponential model led

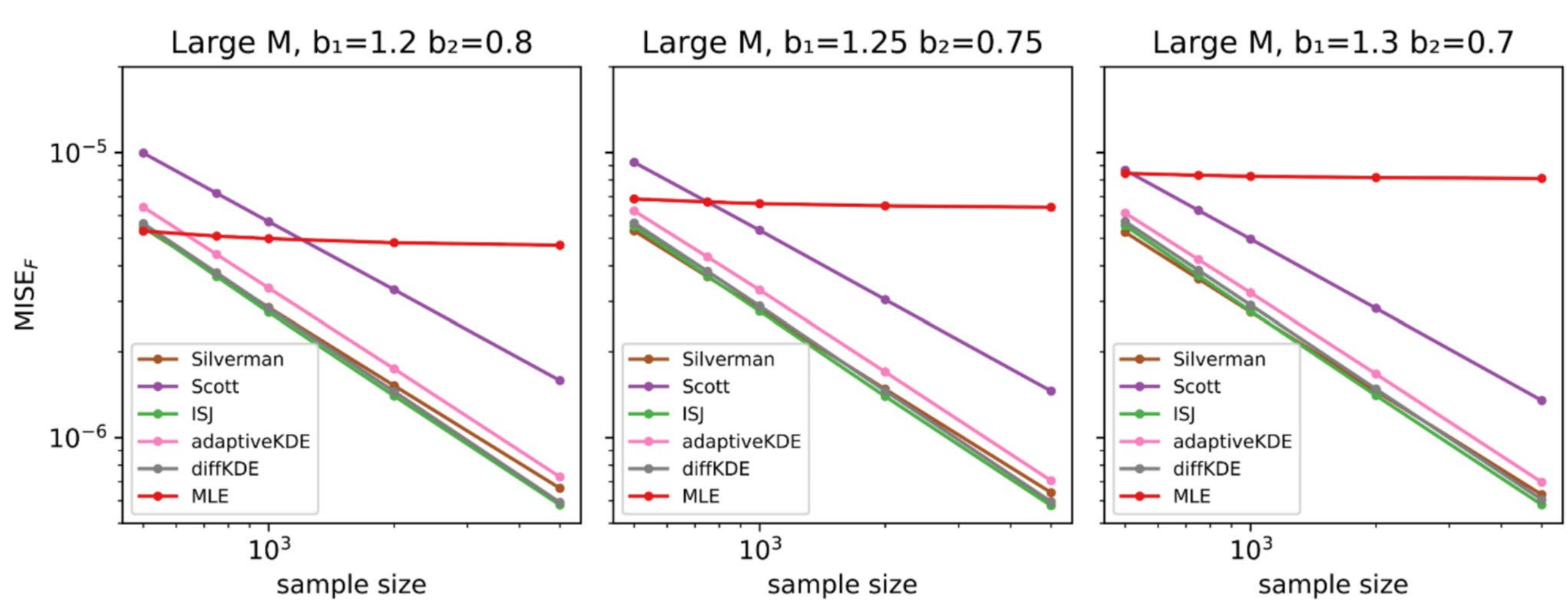

**Fig. 6** $\mathrm{MISE}_F$ results for the data drawn from the concave bi-exponential distribution (10–11) with $b_1 = 1.2$–left, $b_1 = 1.25$–middle, and $b_1 = 1.3$–right, where $b_2 = 1 - b_1$. Colors mark the KDE methods of CDF estimation and the MLE estimate of CDF assuming the exponential distribution model (red)

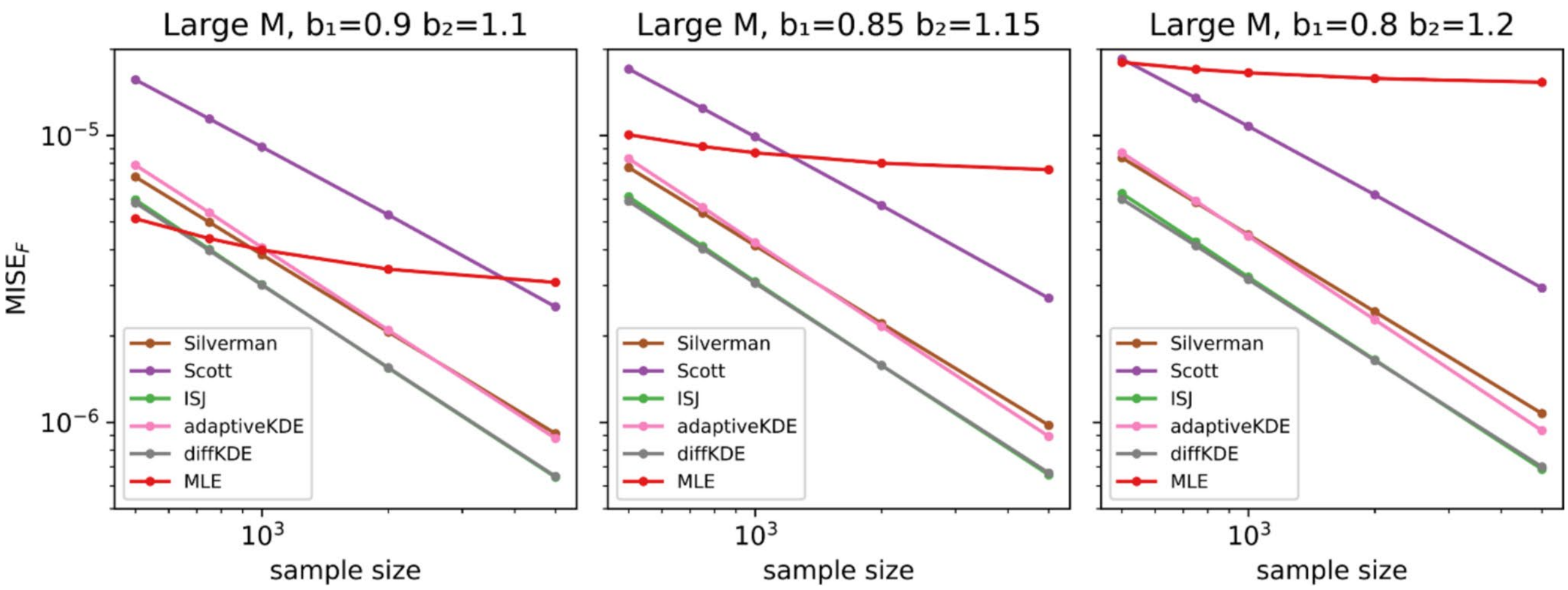

**Fig. 7** $MISE_F$ results for the data drawn from the convex bi-exponential distributions (10–11) with $b_1=0.9$–left, $b_1=0.85$–middle, and $b_1=0.8$ – right, where $b_2=1-b_1$. Colors mark the KDE methods of CDF estimation and the MLE estimate of CDF assuming the exponential distribution model (red)

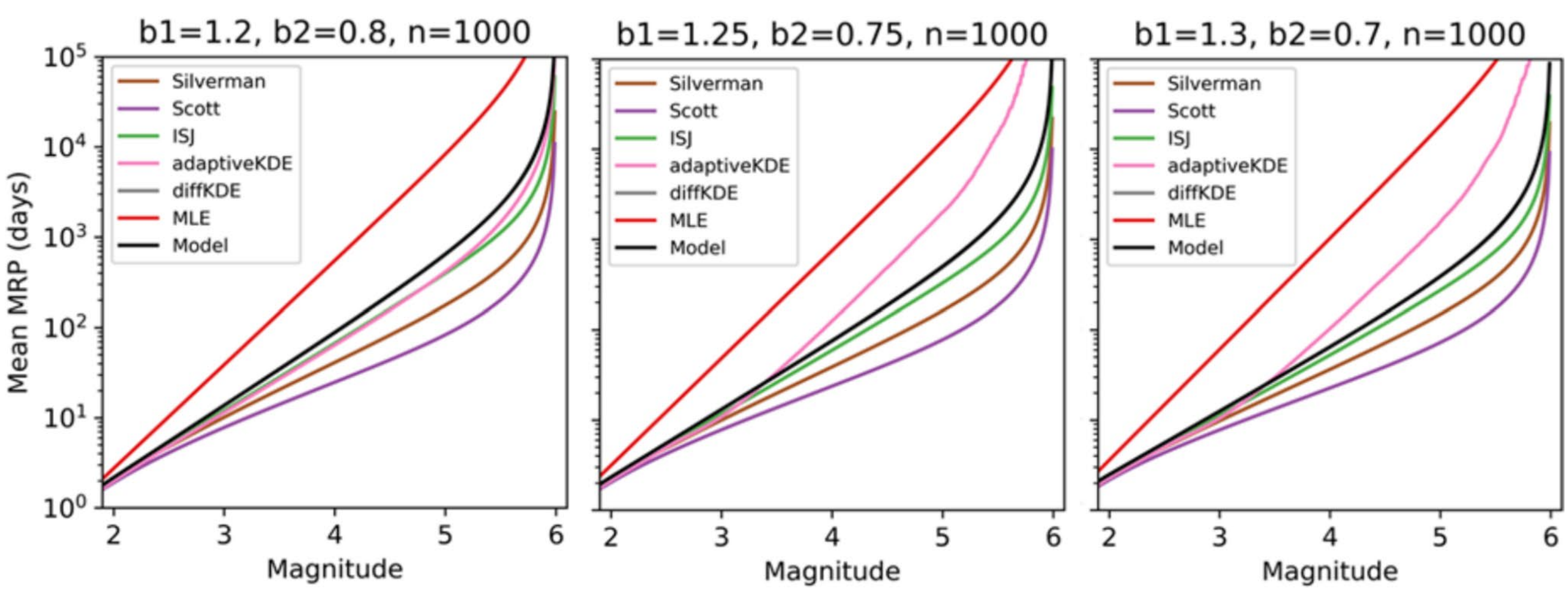

**Fig. 8** Mean MRP using 1000 magnitude values drawn from the concave bi-exponential distribution (10–11) with $b_1=1.2$–left, $b_1=1.25$–middle, and $b_1=1.3$–right, where $b_2=1-b_1$. Colors mark MRPs for CDFs obtained from the studied KDE methods, for CDF from the MLE assuming the exponential distribution model (MLE—red), and for the CDF of the data distribution (Model—black)

to dramatic, unacceptable discrepancies between the estimated and true MRPs. For example, when the underlying distribution was the concave bi-exponential distributions with $b_1=1.3$ and $b_2=0.7$, the true MRP for a magnitude 4 event was around 60 days compared to the MLE estimate of around 1000 days.

### Exponential-Gaussian distribution

The results obtained using the exponential-Gaussian distribution are shown in Figs. 10 ($MISE_F$) and 11 (MRP). Here, the performance of all KDE methods, except Scott's method, was comparable. For samples greater than 3000 events, the adaptiveKDE performed the best, better than the diffKDE, yet the differences in $MISE_F$ were minor. In the 1000-element sample case shown in Fig. 11, the MRP estimated from diffKDE nearly overlaps the true MRP (model). The MRP estimates from the diffKDE method were the most accurate of the five KDE methods tested.

## Conclusions

Of the five KDE methods tested, diffKDE provides the most accurate estimate of the cumulative distribution function of magnitude in the range of large magnitudes. Since large

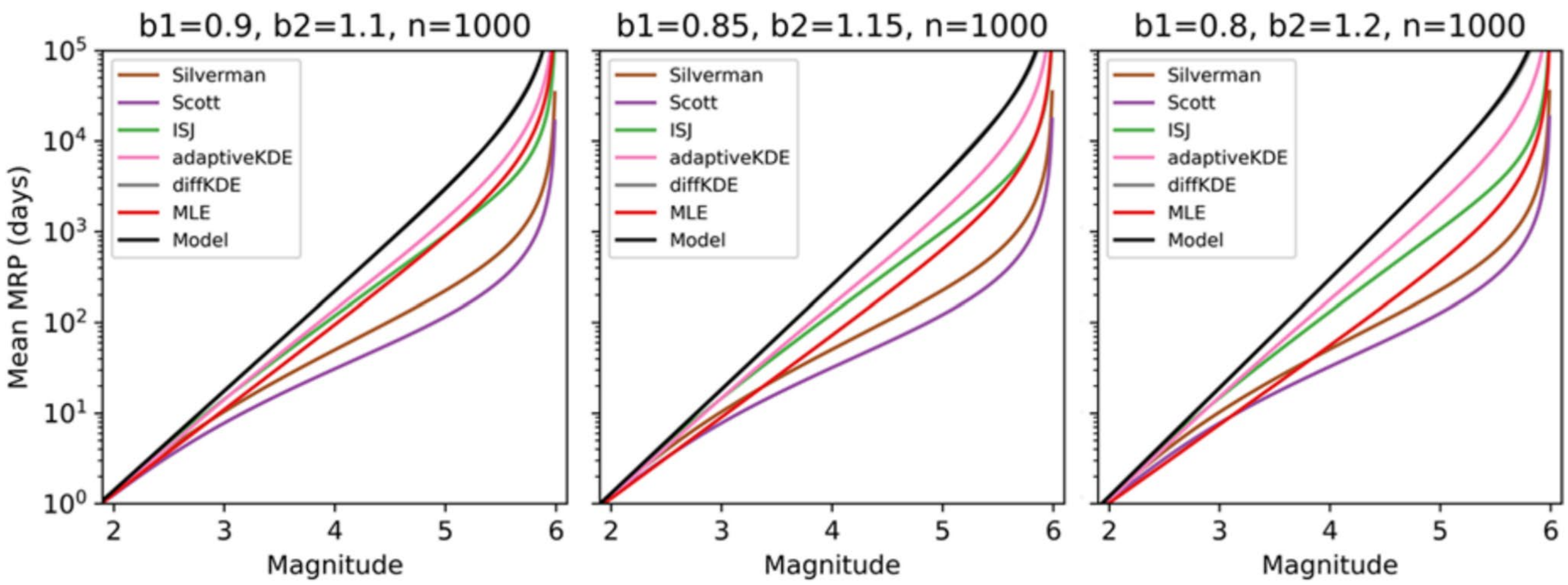

**Fig. 9** Means MRP using 1000 magnitude values drawn from the convex bi-exponential distributions (10–11) with $b_1=0.9$–left, $b_1=0.85$–middle, and $b_1=0.8$–right, where $b_2=1-b_1$. Colors mark MRPs for CDFs obtained from the studied KDE methods, for CDF from the MLE assuming the exponential distribution model (MLE—red), and for the CDF of data distribution (Model—black)

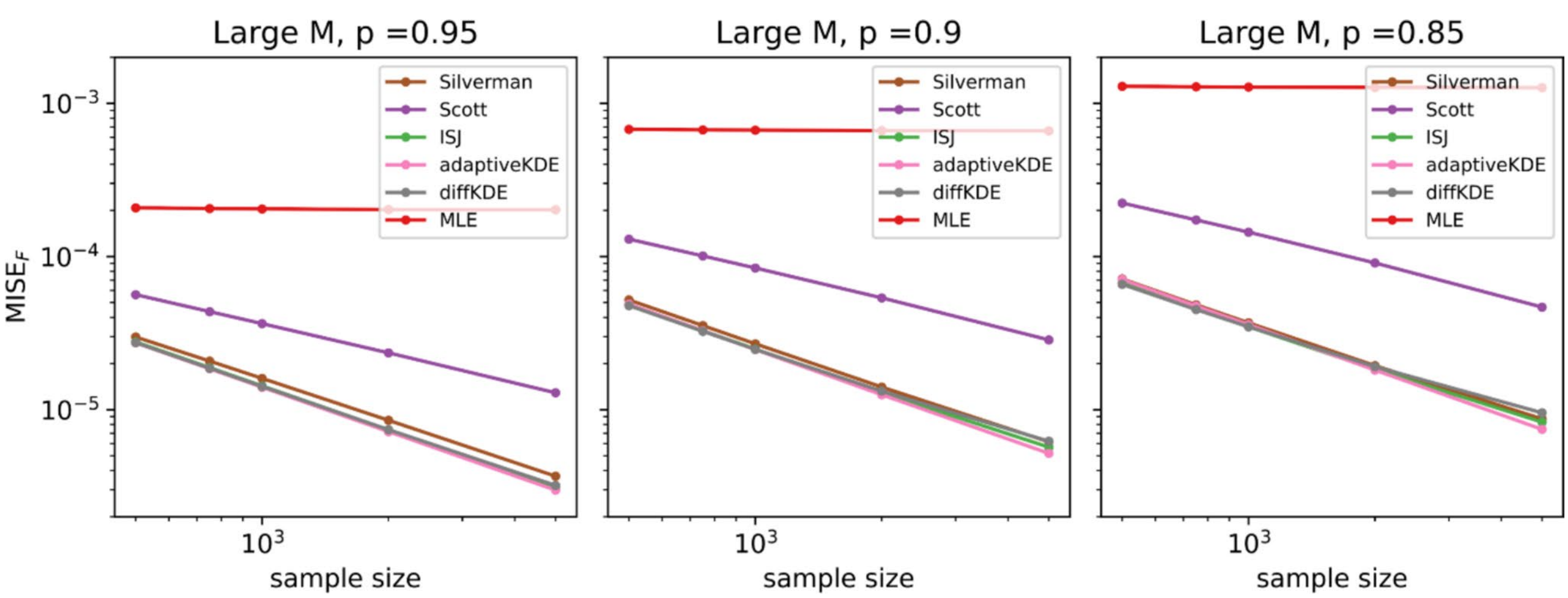

**Fig. 10** $MISE_F$ results for the data drawn from the exponential-Gaussian distribution (Eq. 12) with $p=0.95$–left, $p=0.9$–middle, and $p=0.85$–right. Colors mark the KDE methods of CDF estimation and the MLE estimate of CDF assuming the exponential distribution model (red)

magnitude events are of primary concern in seismic hazard analysis, we conclude that diffKDE is the most suited for probabilistic seismic hazard assessment out of these five methods. When there are enough data samples, 400 events or more—a condition not unusual in anthropogenic seismicity—the estimates using the diffKDE method agree well with actual hazard values.

Obviously, when the underlying distribution is exponential, the MLE method based on the exponential distribution model provides excellent hazard estimates. Nevertheless, our simulations showed that the diffKDE estimates are also acceptable in such a case; hence, the diffKDE method can be safely used even if the magnitude distribution is exponential.

When the distribution that underlies the data deviates from exponentiality, which is an often case in anthropogenic seismicity (Kostoglou et al. 2025), the use of the exponential model of magnitude distribution as a base for hazard estimation results in very inaccurate hazard estimates, whose practical consequences can be severe. Therefore, using the diffKDE estimation method is recommended in the probabilistic analysis of anthropogenic seismic hazard unless there is direct evidence from a goodness-of-fit test that the magnitude distribution is exponential.

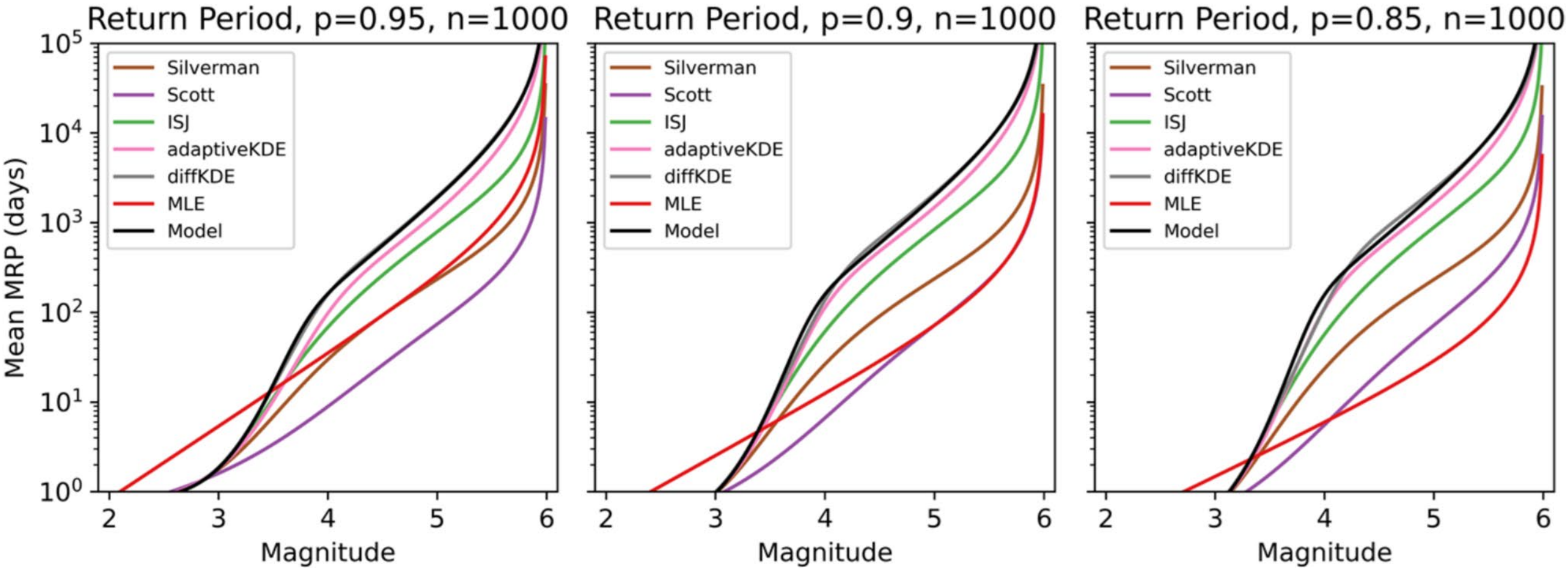

**Fig. 11** Mean MRP for the 1000 magnitude values drawn from the exponential-Gaussian distribution (Eq. 12) with $p=0.95$–left, $p=0.9$–middle, and $p=0.85$–right. Colors mark MRPs for CDFs obtained from the studied KDE methods, for CDF from the MLE assuming the exponential distribution model (MLE—red), and for the CDF of the data distribution (Model—black)

## Appendix—MISE$_F$ across all magnitudes

See Figs. 12, 13, 14, and 15

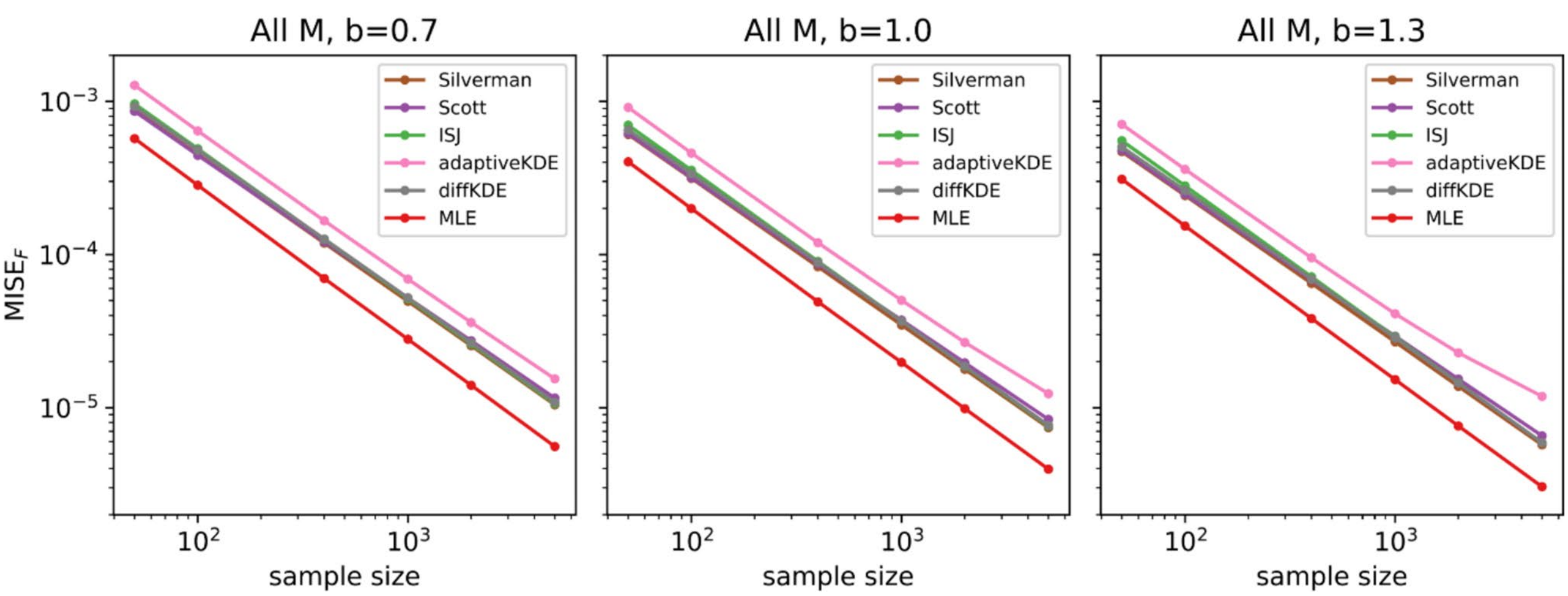

**Fig. 12** Exponential distribution. MISE$_F$ for the full magnitude range of 0.5–6

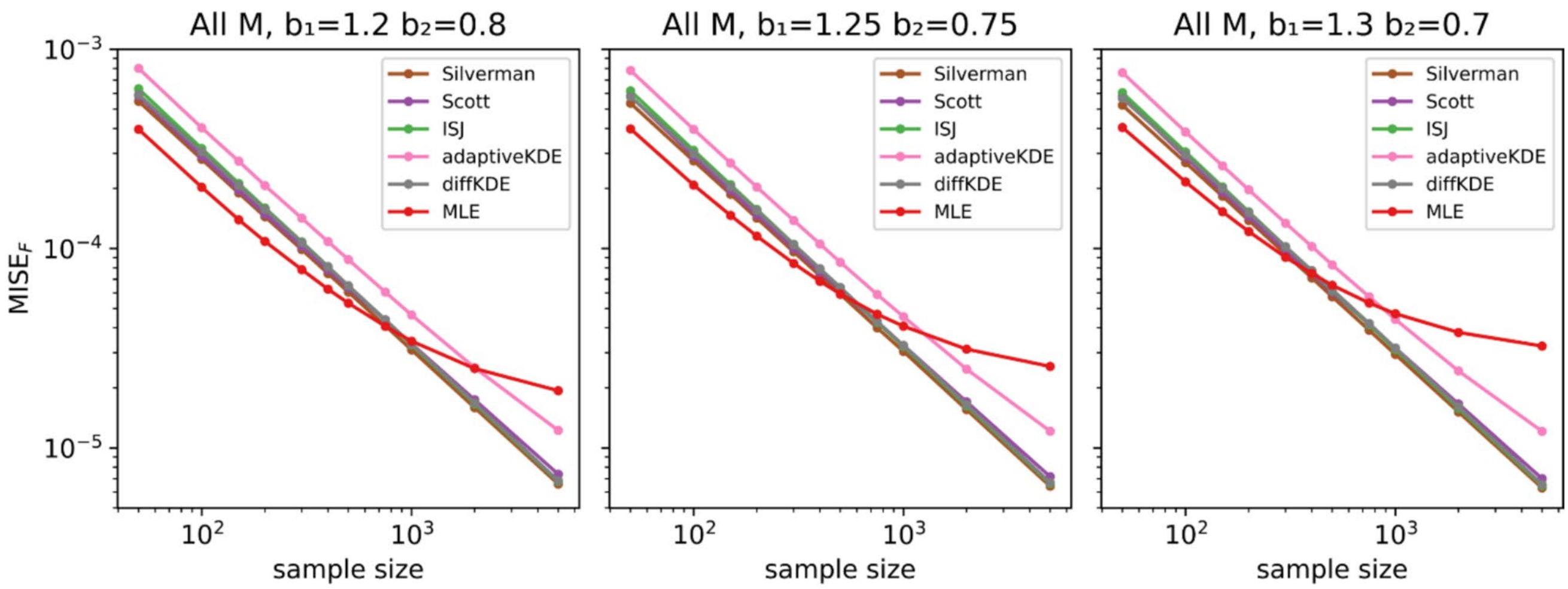

**Fig. 13** Bi-exponential distribution–concave case. $MISE_F$ for the full magnitude range of 0.5–6

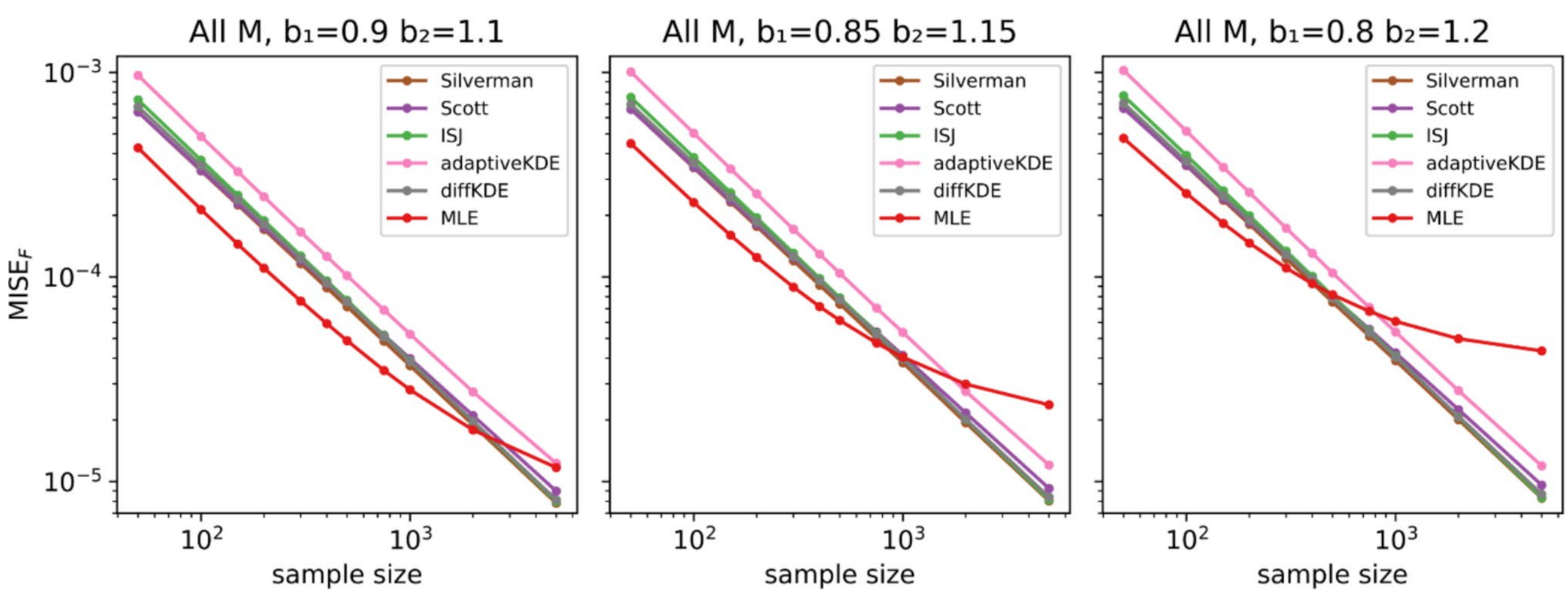

**Fig. 14** Bi-exponential distribution–convex case. $MISE_F$ for the full magnitude range of 0.5–6

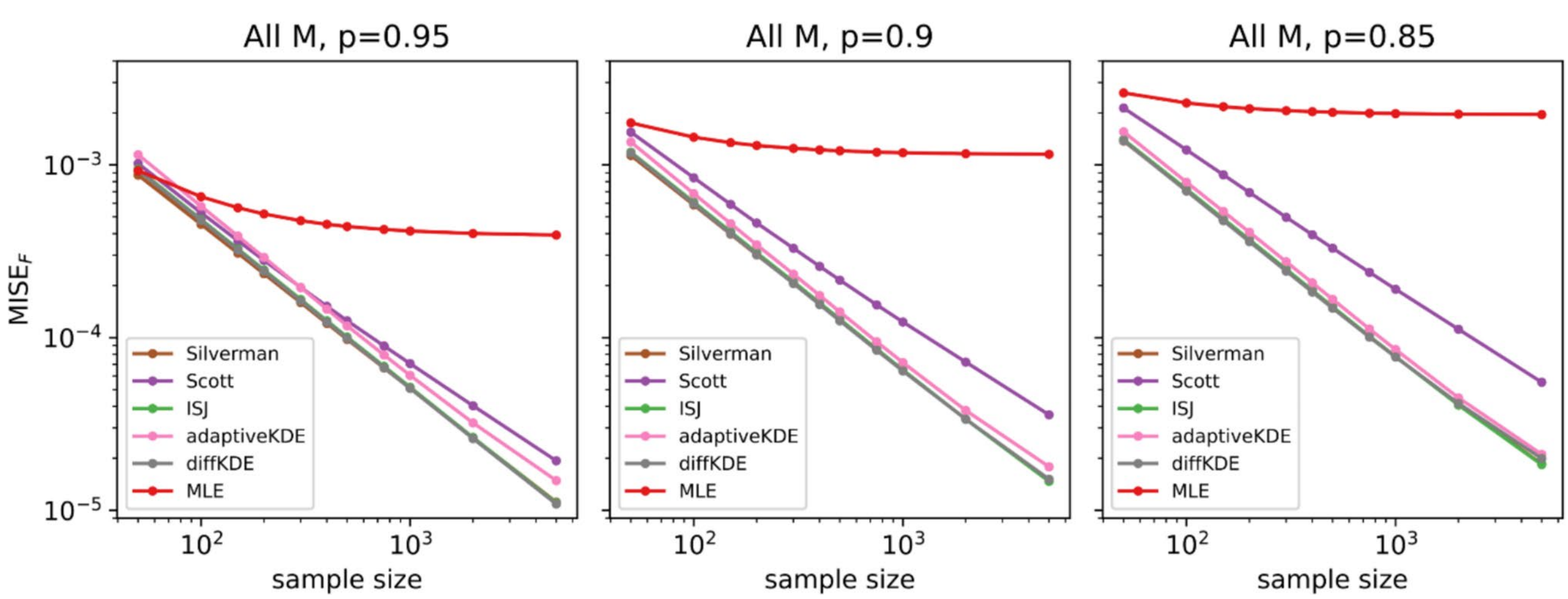

**Fig. 15** Exponential-Gaussian distribution. $MISE_F$ for the full magnitude range of 0.5–6

**Acknowledgements** We acknowledge the CINECA award under the ISCRA (Italian SuperComputing Resource Allocation) initiative, for the availability of high-performance computing resources and support. The work presented in this paper was supported by the project DT-Geo (A Digital Twin for GEOphysical extremes) funded by Horizon Europe under the grant agreement No 101058129.

**Author contribution** Francis Tong contributed to writing—original draft preparation, investigation, methodology, and software. Stanisław Lasocki contributed to validation, methodology, supervision, and writing—review and editing. Beata Orlecka-Sikora contributed to conceptualization, methodology, and funding acquisition.

**Funding** Francis Tong and Beata Orlecka-Sikora report that financial support was provided by Horizon Europe grant DT-Geo 101058129. Stanislaw Lasocki reports that financial support was provided by the Ministry of Education and Science of the Republic of Poland.

## Declarations

**Conflict of interest** The authors declare the following financial interests/personal relationships which may be considered as potential competing interests: Francis Tong reports financial support was provided by Horizon Europe. Stanislaw Lasocki reports financial support was provided by the Ministry of Education and Science of the Republic of Poland. Beata Orlecka-Sikora reports financial support was provided by Horizon Europe.

**Code availability** The code used in this study can be found at https://github.com/FrancisTong/kdetest.